\documentclass[prd,reprint,superscriptaddress,preprintnumbers,eqsecnum,showpacs,nofootinbib,nobibnotes]{revtex4-1}

\usepackage{amsfonts,amsmath,amssymb,bm,natbib,color}
\usepackage{graphicx}
\usepackage{natbib} 

\newcommand{\be}{\begin{equation}}
\newcommand{\bea}{\begin{eqnarray}}
\newcommand{\ee}{\end{equation}}
\newcommand{\eea}{\end{eqnarray}}

\def\1eq#1{Eq.~(\ref{#1})}

\def\2eqs#1#2{Eqs.~(\ref{#1}) and~(\ref{#2})}
\def\3eqs#1#2#3{Eqs.~(\ref{#1}),~(\ref{#2}) and~(\ref{#3})}

\def\kint{\int_k\!}
\def\diff{{\rm d}}

\def\g{\widetilde\Gamma^{\mathrm{np}}}

\def\hh{m^2}

\def\Cgh{\widetilde{C}_{\mathrm{gh}}}
\def\Cgl{\widetilde{C}_{\mathrm{gl}}}

\def\fgl{f_{\mathrm{gl}}}
\def\fgh{f_{\mathrm{gh}}}

\begin{document}

\title{Effects of the ghost sector in gluon mass dynamics}

\author{A.~C. Aguilar}
\affiliation{University of Campinas - UNICAMP, 
Institute of Physics ``Gleb Wataghin'',
13083-859\\ Campinas, SP, Brazil}

\author{C.~T. Figueiredo}
\email{Poster presented  by C. T. Figueiredo at the ``International Workshop on Hadron Physics'', 18-23 March 2018, Florian\'opolis, Brazil.}
\affiliation{University of Campinas - UNICAMP, 
Institute of Physics ``Gleb Wataghin'',
13083-859\\ Campinas, SP, Brazil}

\begin{abstract}

In this work we investigate the effects of the ghost sector on the dynamical mass generation for the gauge boson of a pure Yang-Mills theory.
The generation of a dynamical mass for the gluon is realized by the Schwinger mechanism, 
which is triggered by the existence of longitudinally coupled massless poles 
in the fundamental vertices of the theory.
The appearance of such poles occurs by purely dynamical reasons and 
is governed by a set of Bethe-Salpeter equations.
In previous studies, only the presence of massless poles in the 
background-gauge three-gluon vertex was considered. 
Here, we include the possibility for such poles to appear also in the corresponding ghost-gluon vertex.
Then, we solve the resulting Bethe-Salpeter system, 
which reveals that the contribution associated with the poles of the ghost-gluon vertex is suppressed
with respect to those originating from the three-gluon vertex.

\end{abstract}


\maketitle

\section{\label{sec:intro} Introduction}   

In the last decade, the nonperturbative generation of a dynamical gluon mass has attracted notable attention~\mbox{\cite{Cloet:2013jya,Aguilar:2015bud,Roberts:2016vyn}}. 
In particular, lattice results reveal that the gluon propagator, as well as the ghost dressing function, 
remain finite in the infrared region of QCD~\cite{Cucchieri:2007md,Cucchieri:2007rg,Cucchieri:2009zt,Bowman:2007du,Bogolubsky:2009dc,Oliveira:2009eh,Ayala:2012pb,Bicudo:2015rma}, which has been interpreted as a consequence of an effective gluon mass~\cite{Cornwall:1981zr,Aguilar:2006gr,Aguilar:2008xm, Aguilar:2011xe,Ibanez:2012zk,Aguilar:2016vin}.

In this work, we use a synthesis of the Background Field Method and the Pinch Technique formalisms, 
known in the literature as PT-BFM~\cite{Binosi:2003rr,Aguilar:2006gr,Aguilar:2008xm,Binosi:2007pi,Binosi:2008qk}, to study the phenomenon of a dynamical mass generation for the gluon. 

In this scheme, the gluon fields $(A^{a}_{\mu})$ are described as the sum of a
quantum $(Q^{a}_{\mu})$ and a background $(B^{a}_{\mu})$ part. 
The quantum part $(Q)$ behaves as the conventional QCD gluon, while the 
background $(B)$ behaves as an Abelian field. This separation 
introduces mixed Green's functions, describing combinations of $B$
and $Q$ fields. For instance, we have three types of gluon propagators: (i)
the conventional $(Q^2)$ propagator, $\Delta^{\mu\nu}(q)$, formed by
contraction of two $Q$ gluons, (ii) the quantum-background $(QB)$ propagator,
$\widetilde{\Delta}^{\mu\nu}(q)$, with one $Q$ and one $B$ gluon, and (iii)  the background $(B^2)$ propagator, 
$\widehat{\Delta}^{\mu\nu}(q)$ with two $B$-type gluons.

In order to obtain a massive solution for the gluon propagator, $\Delta^{\mu\nu}(q)$, without breaking the gauge symmetry of the theory, 
we need to invoke the well-known Schwinger mechanism~\cite{Schwinger:1962tn,Schwinger:1962tp}. 
Within the PT-BFM scheme, the Schwinger mechanism is integrated to the gluon propagator Schwinger-Dyson equation 
(SDE) through its vertices, which must contain 
longitudinally dynamical massless poles of the 
generic form $q^{\mu}/q^2 \widetilde{C}(q,r,p)$~\cite{Aguilar:2006gr,Aguilar:2007ie,Aguilar:2008xm,Aguilar:2017dco}.

In QCD the appearance of these poles can occur by purely dynamical reasons, where the formation of the (colored) massless bound states requires sufficiently strong binding couplings~\cite{Jackiw:1973tr,Cornwall:1973ts,Jackiw:1973ha}.
 
Recently, an approximated description for the formation of such 
massless poles in the structure of the three-gluon vertex 
composed of one background gluon and two quantum gluons ($BQ^2$) was obtained~\cite{Aguilar:2011xe}.  
In this approximation, only the one-loop dressed gluonic diagram was considered in the Bethe-Salpeter equation (BSE) which controls the dynamics of the three-gluon vertex. 
In this presentation, we will include the contribution of poles in the ghost-gluon vertex  with a background gluon ($Bc\bar{c}$) as well, 
in order to investigate the effects of the ghost sector in gluon mass dynamics~\cite{Aguilar:2017dco}.

\section{\label{sec:gluon}  Gluon mass and vertices with massless poles}

In the Landau gauge, we can write the $(Q^2)$ gluon propagator as 
\begin{align}
	\Delta_{\mu\nu}(q) &= -i\Delta(q^2)P_{\mu\nu}(q); \;\; P_{\mu\nu}(q) = g_{\mu\nu}-\frac{q_\mu q_\nu}{q^2},
\end{align}	
where $\Delta(q^2)$ represents the scalar part of the gluon propagator and obeys 
$\Delta^{-1}(q^2) = q^2 + i \Pi (q^2)$, with $\Pi (q^2)$ being 
the scalar form factor of the gluon self-energy  \mbox{$\Pi_{\mu\nu}(q) = P_{\mu\nu}(q) \Pi (q^2)$}.
Additionally, the ghost propagator is given by
$D(q^2)=iF(q^2)/q^2$, where $F(q^2)$ is the so-called ghost dressing function.

Within the PT-BFM framework, the SDE for the gluon propagator is expressed in terms of 
the special $QB$ gluon self-energy $\widetilde\Pi_{\mu\nu}(q)$, so that  
\begin{align}
	\Delta^{-1}(q^2)P_{\mu\nu}(q) &= \frac{q^2P_{\mu\nu}(q)  + i
	\widetilde{\Pi}_{\mu\nu}(q)}{1 + G(q^2)}, 
\label{glSDE1}
\end{align}
where $G(q^2)$ is the``ghost-gluon mixing self-energy'', which plays a key role in the pinch-technique~\cite{Binosi:2009qm,Aguilar:2008xm,Binosi:2014aea}. In addition, in  the Landau gauge, it coincides with the inverse of the  ghost dressing function at zero momentum {\it i.e.} \mbox{$1+G(0) = F^{-1}(0)$}~\cite{Aguilar:2009nf}.
The advantage of expressing such SDE in terms of $\widetilde\Pi_{\mu\nu}(q)$, 
instead of the conventional self-energy $\Pi_{\mu\nu}(q)$, 
is that, in doing so, each vertex,
when contracted with the momentum carried by $B$-gluon,
will satisfy an Abelian-like Slavnov-Taylor identity (STI). 
Specifically, the $BQ^2$ 
vertex, $\widetilde{\Gamma}_{\mu\alpha\beta}$, and the $Bc\bar c$ 
vertex, $\widetilde{\Gamma}_{\mu}$, obey (color omitted and all momenta entering)
\begin{align}
	&q^\mu \widetilde{\Gamma}_{\mu\alpha\beta}(q,r,p) = i\Delta_{\alpha\beta}^{-1}(r) - i\Delta_{\alpha\beta}^{-1}(p), \nonumber\\
	&q^\mu \widetilde{\Gamma}_\mu(q,r,p) = iD^{-1}(r^2) - iD^{-1}(p^2).\label{AbWI3gh}
\end{align}

Assuming there are no massless poles in these vertices, 
we can use the Taylor expansion of both sides of the equations above, 
in order to generate the corresponding Ward-Takahashi identities (WTIs) which is valid in the limit of $q \to 0$
\begin{align}
    &\widetilde{\Gamma}_{\mu\alpha\beta}(0,r,-r) = -i \frac{\partial }{\partial r^\mu}\Delta^{-1}_{\alpha\beta}(r), \nonumber \\
    &\widetilde{\Gamma}_{\mu}(0,r,-r) = -i \frac{\partial }{\partial r^\mu}D^{-1}(r^2).
    \label{WTI3gh}	
\end{align}
	
Recently, it was demonstrated that, if the 
PT-BFM vertices with a $B$ leg of momentum $q$ 
do not contain massless poles of the type 
$1/q^2$, then the inverse gluon propagator $\Delta^{-1}(q^2)$, given in \1eq{glSDE1}, is rigorously  zero, 
so that, the gluon remains massless~\cite{Aguilar:2016vin}. 
The demonstration benefits from an integral relation, valid in dimensional regularization,  known as the ``seagull identity''~\cite{Aguilar:2016vin}. Then, it is possible to show that, in the absence of $1/q^2$ poles\footnote{We  have introduced the compact notation where \mbox{$\int_{k}\equiv\mu^{\epsilon}/(2\pi)^{d}\!\int\!\mathrm{d}^d k$}, with $d=4-\epsilon$ the space-time dimension, and $\mu$  the 't Hooft mass scale.} 
\begin{align}
    \Delta^{-1}(0) & = \underbrace{\int_k\frac{\partial}{\partial k_\mu}{\cal F}_\mu(k) =0}_{\rm seagull \,\, identity},
    \label{seag1}
\end{align}
where ${\cal F}_\mu(k)=k_\mu{\cal F}(k^2)$, with ${\cal F}(k^2)$ being an arbitrary scalar function,
which vanishes rapidly enough as \mbox{$k^2\rightarrow\infty$}~\cite{Aguilar:2016vin}.

\begin{figure}[t]
\includegraphics[scale=0.32]{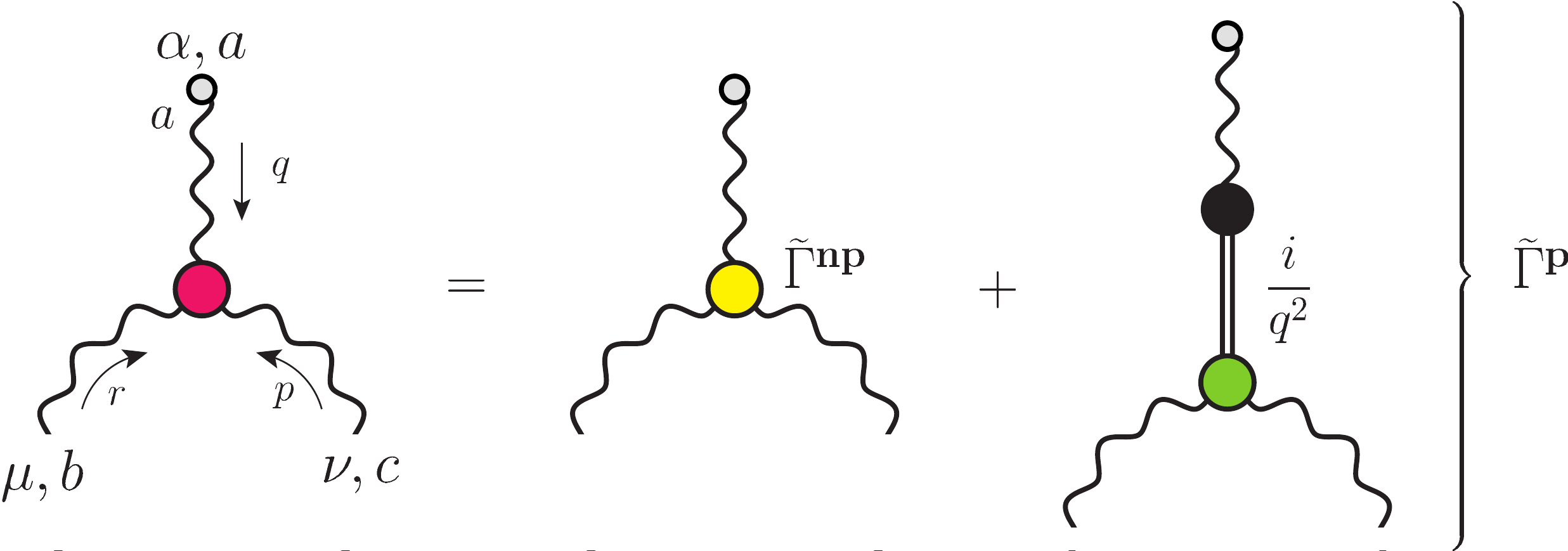}
\vspace{-0.3cm}
\caption{Three-gluon vertex $(BQ^2)$ divided into one part that does not contain pole in $q^2$ and another that does.  Analogous decomposition holds for the ghost-gluon vertex ($Bc\bar{c}$).}
\label{fig:pole}
\end{figure}

The cancellation of Eq.~(\ref{seag1}) can be evaded by introducing longitudinally coupled $1/q^2$ poles 
to the PT-BFM vertices cited above. For example, in Fig.~\ref{fig:pole}, we illustrate the division of the three-gluon $BQ^2$ vertex into one part that does not contain pole in $q^2$ (yellow) and another that does (green). 
 Similar decomposition holds for the ghost-gluon vertex, $Bc\bar{c}$.
As we will see in what follows, such inclusion triggers the Schwinger mechanism, 
allowing the generation of a gauge boson mass. 

In this presentation, we will consider the possibility of poles for both $BQ^2$ and $Bc\bar c$ vertices, then one has
\begin{align}
    \widetilde{\Gamma}_{\mu\alpha\beta}(q,r,p) &= \g_{\mu\alpha\beta}(q,r,p) + i \frac{q_\mu}{q^2}\widetilde{C}_{\alpha\beta}(q,r,p),
    \nonumber\\ 
    \widetilde{\Gamma}_{\mu}(q,r,p)&=\g_{\mu}(q,r,p)+ i \frac{q_\mu}{q^2}\Cgh(q,r,p), 
    \label{GnpGp2}
\end{align}
where the superscript ``np'' stands for ``no-pole'' and $\widetilde{C}_{\alpha\beta}$ and $\Cgh$ are the bound-state gluon-gluon and gluon-ghost wave functions, respectively.
%
Then, to keep the symmetry of the theory intact, we require that the STIs of \eqref{AbWI3gh} preserve their form when including the poles, thus  
\begin{align}
	& q^\mu \g_{\mu\alpha\beta}+ \widetilde{C}_{\alpha\beta} = i\Delta_{\alpha\beta}^{-1}(r) - i\Delta_{\alpha\beta}^{-1}(p), \nonumber \\
    & q^\mu\g_{\mu} +\Cgh = iD^{-1}(r^2) - iD^{-1}(p^2). 
    \label{ghSTIwP}
\end{align}
where the vertices and the bound state wave functions
are functions of $(q,r,p)$. We can now take the limit as $q\to 0$, so that the zeroth order terms in $q$ yields 
\begin{align}
    \widetilde{C}_{\alpha\beta}(0,r,-r)&=0;&
    \Cgh(0,r,-r)&=0,
    \label{zerothC}
\end{align}
while the terms linear in $q$ provide a new set of WTIs, 
\begin{align}
	\g_{\mu\alpha\beta} &= -i\frac{\partial}{\partial r^\mu}\Delta^{-1}_{\alpha\beta}(r)
	- \left\lbrace\frac{\partial}{\partial q^\mu}\widetilde{C}_{\alpha\beta}(q,r,-r-q)\right\rbrace_{q=0},& \nonumber \\  
   \g_\mu &= -i\frac{\partial}{\partial r^\mu} D^{-1}(r^2)- \left\lbrace\frac{\partial}{\partial q^\mu}\Cgh(q,r,-r-q)\right\rbrace_{q=0}.
    \label{WI2ghwithpole}
\end{align}

The first terms on the r.h.s. of Eq.~(\ref{WI2ghwithpole}) lead to a result in the form of \1eq{seag1}, so their contributions vanish. However, 
the second terms survive, from which we obtain~\cite{Aguilar:2017dco}.
\begin{align}
\Delta^{-1}(0) &=\lambda \int_k k^2 \Delta^2(k^2)\left[1-\frac32g^2C_AY(k^2)\right]\Cgl'(k^2) & \nonumber\\
&-\frac\lambda3\int_k k^2 D^2(k^2)\Cgh'(k^2)\,,
\label{DSEmass}
\end{align}
where $\lambda = 3g^2C_AF(0)/2 $ with $C_A$ being the Casimir eigenvalue of the adjoint representation and $\Cgl$ is the form factor of the metric tensor $g_{\alpha\beta}$ in the tensorial decomposition of $\widetilde{C}_{\alpha\beta}$. 
Additionally, we defined
\begin{align}
    C_i^{\prime}(k^2)=\lim_{q\to0}\left\lbrace\frac{\partial \widetilde{C}_i(q,k,-k-q)}{\partial (k+q)^2}\right\rbrace, \;\;i=\mathrm{gl},\mathrm{gh}.
\end{align}

From \1eq{DSEmass}, we notice that, in order for $\Delta^{-1}(0)$ to acquire a nonvanishing value, we need at least one of $\Cgl^{\prime}$ and $\Cgh^{\prime}$ do not vanish identically.
In addition, it is possible to establish a link between $\Cgl^\prime$ and a running gluon mass through~\cite{Aguilar:2017dco} 
\begin{align}
    \hh(q^2)=\Delta^{-1}(0)+\int_0^{q^2}\!\!\diff y\,\Cgl^\prime(y)\,.
    \label{Cglvsmass}
\end{align}

\section{\label{sec:poles}  Dynamics of massless pole formation}

\begin{figure}[h]
\includegraphics[scale=0.3,angle=270]{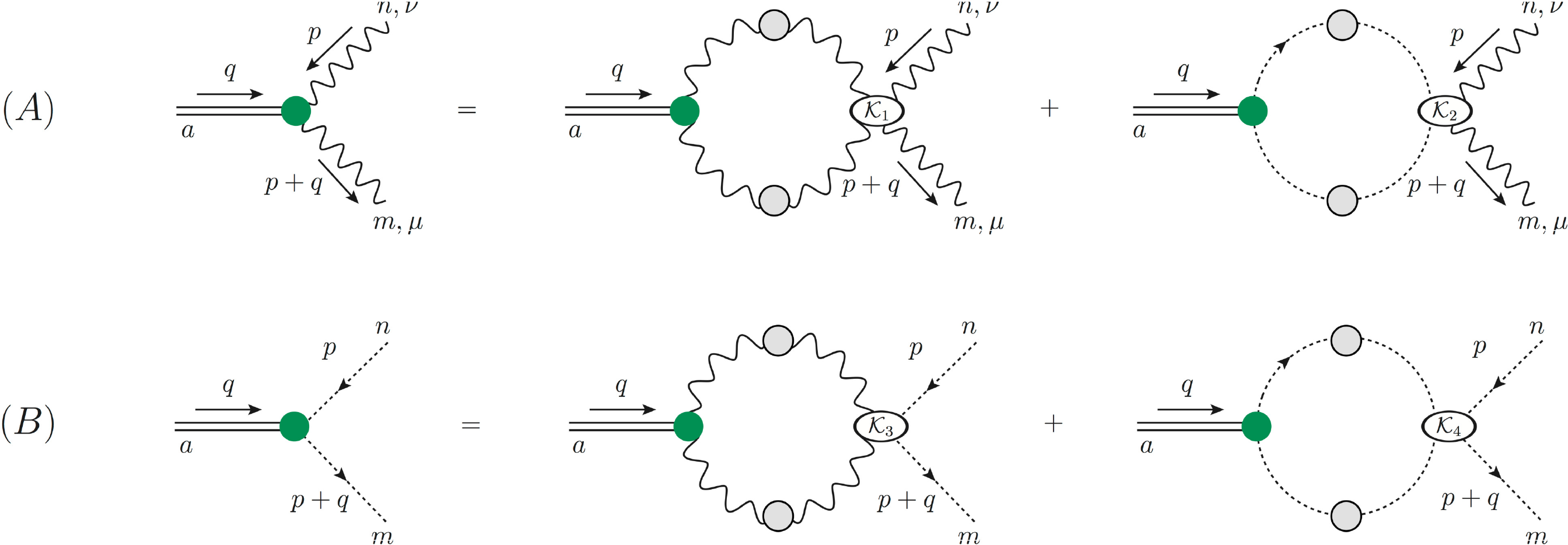}
\caption{{Coupled system of BSEs for the functions $\widetilde{C}_{\mu\nu}$ (top) and  $\Cgh$ (bottom).}}
\label{fig:gluon-ghost-sde}
\end{figure}

From the SDEs satisfied by the $BQ^2$ and $Bc\bar c$ vertices in the limit $q\rightarrow 0$, one can derive a system of integral equations that governs the behavior of  $\Cgl^{\prime}$ and $\Cgh^{\prime}$ as shown
in the Fig.~\ref{fig:gluon-ghost-sde}. From Eq.~(\ref{zerothC}), we know that the zeroth order terms vanish. Thus, the derivative terms will give the leading contributions. To proceed further with the derivation, we approximate the four-point kernels ${\mathcal K}_i$, appearing in the diagrams of the Fig.~\ref{fig:gluon-ghost-sde}.  to their lowest-order set of diagrams. In doing that we arrive at the following 
coupled system of equations~\cite{Aguilar:2017dco} 
%
\vspace{-0.7cm}

\begin{widetext}
\begin{align}
    \Cgl^\prime(q^2)&=\frac{8\pi}3\alpha_sC_A\!\left[\kint\Cgl^\prime(k^2)\Delta^2(k)\Delta(k+q){\cal N}_1(k,q)+\frac14\kint\Cgh^\prime(k^2)D^2(k)D(k+q){\cal N}_2(k,q)\!
    \right]\!,\nonumber \\
    \Cgh^\prime(q^2)&=2\pi\alpha_sC_A\!\left[\kint\Cgl^\prime(k^2)\Delta^2(k)D(k+q){\cal N}_3(k,q)+\frac12\kint\Cgh^\prime(k^2)D^2(k)\Delta(k+q){\cal N}_4(k,q)
    \!\right]\!,
    \label{TheSys}
\end{align}
\end{widetext}
with
\begin{align}
{\cal N}_1(k,q)&=\frac{(q\!\cdot\!k)[q^2k^2-(q\!\cdot\! k)^2]}{q^4k^2(k+q)^2}\fgl^2(k+q)\nonumber \\
 &\hspace{-1.0cm}\times\left[8q^2k^2 + 6(q\!\cdot\!k)(q^2+k^2)+3(q^4+k^4)+(q\!\cdot\! k)^2\right],\nonumber \\
{\cal N}_2(k,q)&=\frac{(q\!\cdot\!k)[q^2k^2-(q\!\cdot\! k)^2]}{q^4}\fgh^2(k+q),\nonumber \\
{\cal N}_3(k,q)&=\frac{(q\!\cdot\!k)[q^2k^2-(q\!\cdot\! k)^2]}{q^2k^2}\fgh^2(k+q),\nonumber \\
{\cal N}_4(k,q)&=\frac{(q\!\cdot\!k)[q^2k^2-(q\!\cdot\! k)^2]}{q^2(k+q)^2}\fgh^2(k+q),
\end{align}
where the $\fgl(r)$ and $\fgh(r)$ are {Ans\"atze} employed for the three-gluon and ghost-gluon vertices. More
specifically, 
\begin{align}
    \Gamma_{\mu\alpha\beta}(q,r,p)=
	\fgl (r)\Gamma^{(0)}_{\mu\alpha\beta}(q,r,p),\nonumber\\
	\Gamma_{\mu}(q,r,p)=
	\fgh (r)\Gamma^{(0)}_{\mu}(q,r,p),
\end{align}
where $\Gamma^{(0)}$ is the tree-level expression for the vertices.

It is interesting to notice that, when we take the limit of \mbox{$q\to0$} in the Eq.~\eqref{TheSys}, $\Cgl(0)$ saturates to a constant~\cite{Binosi:2017rwj}, whereas the structure of the ${\cal N}_3$ and ${\cal N}_4$ kernels implies that $\Cgh(0)=0$.  

\section{\label{sec:Numerical}  Numerical Analysis}

To solve the BSE system  given by Eq.~\eqref{TheSys}, we have to specify 
the following four external functions: $\Delta(q)$, $F(q)$, and
the form factors $\fgl(q)$ and $\fgh(q)$.  For the propagators, 
we use fits for the $SU(3)$ lattice data of the Ref.~\cite{Bogolubsky:2009dc},
whereas for the form factors, we employ  their expected nonperturbative behavior, in the symmetric configuration, derived either in the lattice or SDE analysis~\cite{Athenodorou:2016oyh,Boucaud:2017obn,Binosi:2017rwj,Aguilar:2013xqa}.

\begin{figure*}[]
\begin{minipage}[b]{0.45\linewidth}
\centering
\includegraphics[scale=0.6]{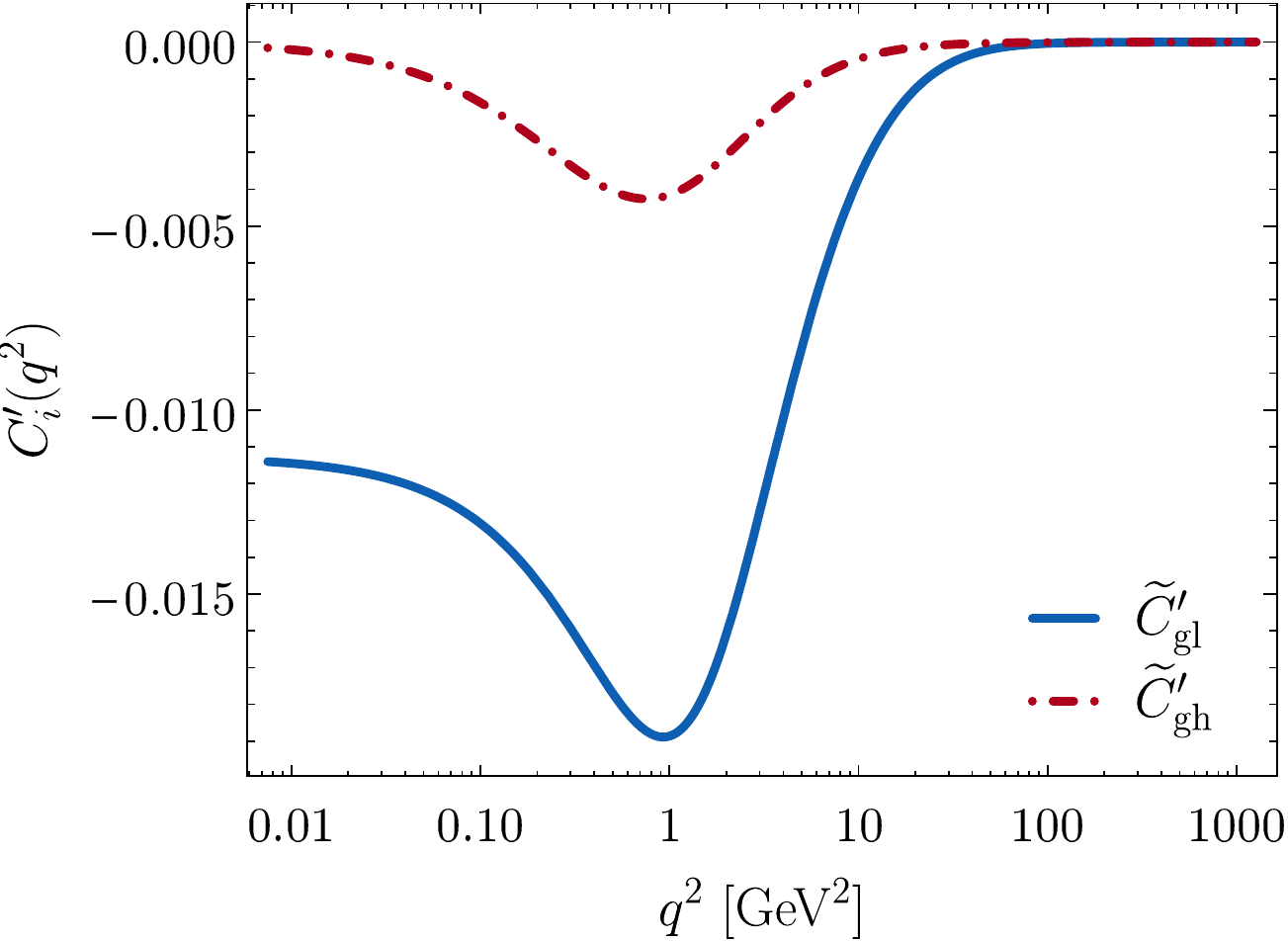}
\end{minipage}
\hspace{1.0cm}
\begin{minipage}[b]{0.45\linewidth}
\includegraphics[scale=0.6]{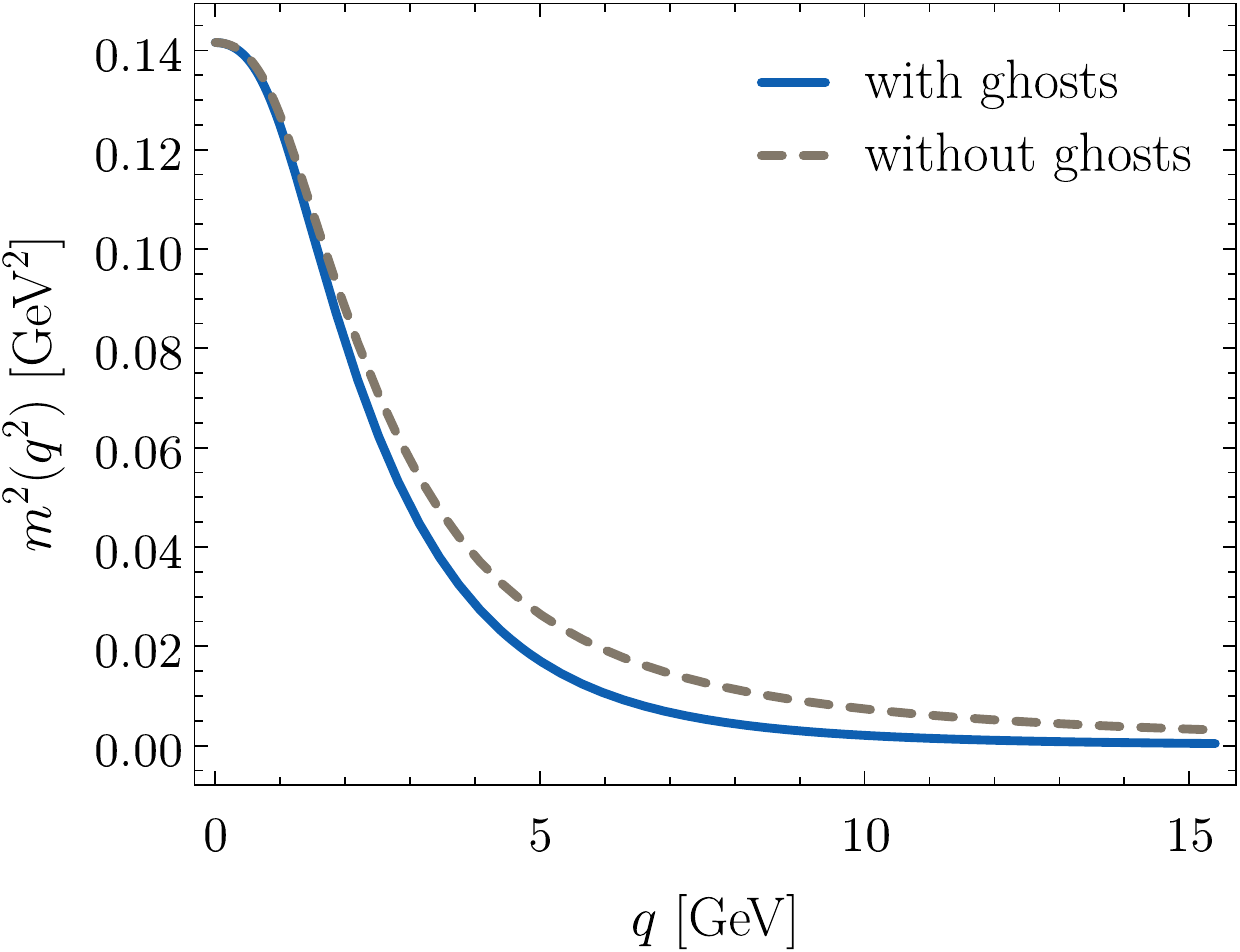}
\end{minipage}
\vspace{-0.3cm}
\caption{\label{fig:C} Normalized gluon and ghost solutions $\Cgl'(q^2)$ and $\Cgh'(q^2)$ of the BSE system
given by Eqs~\eqref{TheSys}. Dynamical gluon mass obtained in this study, compared to the one attained in the absence of ghosts.}
\end{figure*}

Using the quantities specified above, we have solved the coupled system of BSEs~\eqref{TheSys}.
In the left panel of the Fig.~\ref{fig:C}, we show the normalized solution for 
$\Cgl'(q^2)$ and $\Cgh'(q^2)$  obtained for $\alpha_s=0.43$.  In the right  panel, the blue
continuous line represents the resulting gluon mass obtained with Eq.~\eqref{Cglvsmass}, 
while the gray dashed curve is the same quantity obtained in the absence of the 
ghost poles, {\it i.e.} we set in the Eq~\eqref{TheSys} $\Cgh'(q^2)=0$~\cite{Binosi:2017rwj}. From Fig.~\ref{fig:C}, clearly
we notice that the presence of ghosts implies a faster running of the gluon mass. 
Additionally, one can see that the gluon mass,  shown
in the right panel of  Fig.~\ref{fig:C}, can be fitted using the following power-law behavior~\cite{Aguilar:2014tka}
\begin{align}
	m^2(q^2)=m^2(0)/[1+(q^2/m_1^2)^{1+p}], 
\end{align}
where \mbox{$m_1=0.37$ GeV} and \mbox{$p=0.24$} (blue continuous) as opposed to \mbox{$m_1=0.36$ GeV} and \mbox{$p=0.1$}, which represents the case where the effects of the ghosts were
neglected (gray dashed).

\vspace{-0.5cm}

\section{\label{sec:concl}  Conclusions}

Within the PT-BFM scheme, we derived the BSE system which describes the dynamics of massless poles formation 
in the $BQ^2$ and $Bc\bar{c}$ vertices. By solving this system, we were able to obtain non-trivial solutions for both $\Cgl'(q^2)$ and $\Cgh'(q^2)$,
which indicates that the dynamics of QCD is indeed strong enough to generate such poles.

Then, from the numerical results, 
we conclude that the contribution associated with the pole of the gluon-ghost vertex 
is suppressed when compared to that coming from the three-gluon vertex.
The main effect of the presence of ghosts was observed to be 
a slight modification in the running of the gluon mass.  

\vspace{-0.5cm}
\acknowledgments 
\vspace{-0.5cm}

The authors thank the organizers of the {\rm XIV} International Workshop on Hadron Physics for their hospitality. The work of  A.~C.~A  and C. T. Figueiredo  are supported under the 
grants 2017/07595-0 and 2016/11894-0. 
\vspace{-0.5cm}

%

\end{document}